\documentclass{appolb}%
\usepackage{graphicx}
\usepackage{amssymb}
\usepackage{amsmath}
\usepackage{bm}  
\usepackage{amsfonts}

\begin{document}
\renewcommand*{\thefootnote}{\fnsymbol{footnote}}

\title{Doubly heavy $QQ$ tetraquarks\thanks{Presented 
at 
{\em Excited QCD}, Schladming, Austria, January 30 -- February 3, 2019.}}
\author{Michal Praszalowicz
\address{M. Smoluchowski Institute of Physics, Jagiellonian University, \\
ul. S. {\L}ojasiewicza 11, 30-348 Krak{\'o}w, Poland.}}

\maketitle

\begin{abstract}
With the discovery of a doubly charmed $\Xi_{cc}$ baryon a somewhat forgotten issue of
tetraquarks containing two heavy and two light (anti) quarks, ${\cal T}_{QQ}$, triggered theorist's interest.
We discuss quark model estimates of  ${\cal T}_{QQ}$
masses and
a model where the light sector is treated as a soliton. We
show that this model has  different large $N_c$ limit than other approaches.
\end{abstract}


\section{Introduction}

Recent discovery of a doubly charmed $\Xi_{cc}^{++}(3621)$ baryon
by the LHCb Collaboration at CERN \cite{Aaij:2017ueg}
renewed interest in 
$\bar{Q}\bar{Q}q_1q_2$ tetraquarks  or their antiparticles, essentially for
two reasons. Firstly, the LHCb result shows that it is possible to create a $cc$ pair in an attractive channel
that can form a bound state with a light quark. Secondly, on theoretical side, heavy $cc$ pair can be described
to a good 
approximation as a pointlike color  $\overline{\bf 3}$ source
that can form a bound state with two light antiquarks. In the heavy quark limit the $\overline{\bf 3}$ source
acts as a heavy antiquark, and therefore the underlying dynamics is identical to the heavy antibaryon case
(neglecting spin effects). In this paper we shall consider heavy $\bar{Q}\bar{Q}$ pairs acting
as ${\bf 3}$ color source.

Similar excitement arose approximately 17 years ago when SELEX experiment at FermiLab
announced a discovery of $\Xi_{cc}^{+}(3519)$ \cite{Mattson:2002vu}, which is today 
(despite later report 
\cite{Ocherashvili:2004hi})
considered as unconfirmed \cite{PDG}. Phenomenological attempt to estimate the $\bar{c}\bar{c} q_1 q_2$ mass based
on SELEX result can be found {\em e.g.} in Ref.\cite{Gelman:2002wf}, where also large $N_c$ limt
for such states is discussed.

In 1993 Manohar and Wise \cite{Manohar:1992nd} showed 
within heavy quark symmetry approach 
that doubly heavy
tetraquarks are bound in the limit $m_Q \rightarrow \infty$. These arguments were reanalized in 2006
and also very recently, at the time of this conference, by Cohen and collaborators in Refs.~\cite{Cohen:2006jg,Cai:2019orb}.

Asymptotic theorems, however, 
do not provide any hint at what scale they become operative. The aim of
this paper is to recall some simple quark model estimates of the doubly heavy tetraquark mass and then apply a 
phenomenological model
based on a soliton description of the light sector to tetraquarks. At the end we discuss the difference between the soliton
model and regular quark models in the large  $N_c$ limit.

\section{Quark model estimates of the tetraquark mass}

\begin{figure}[h]
\centering
\includegraphics[width=9.1cm]{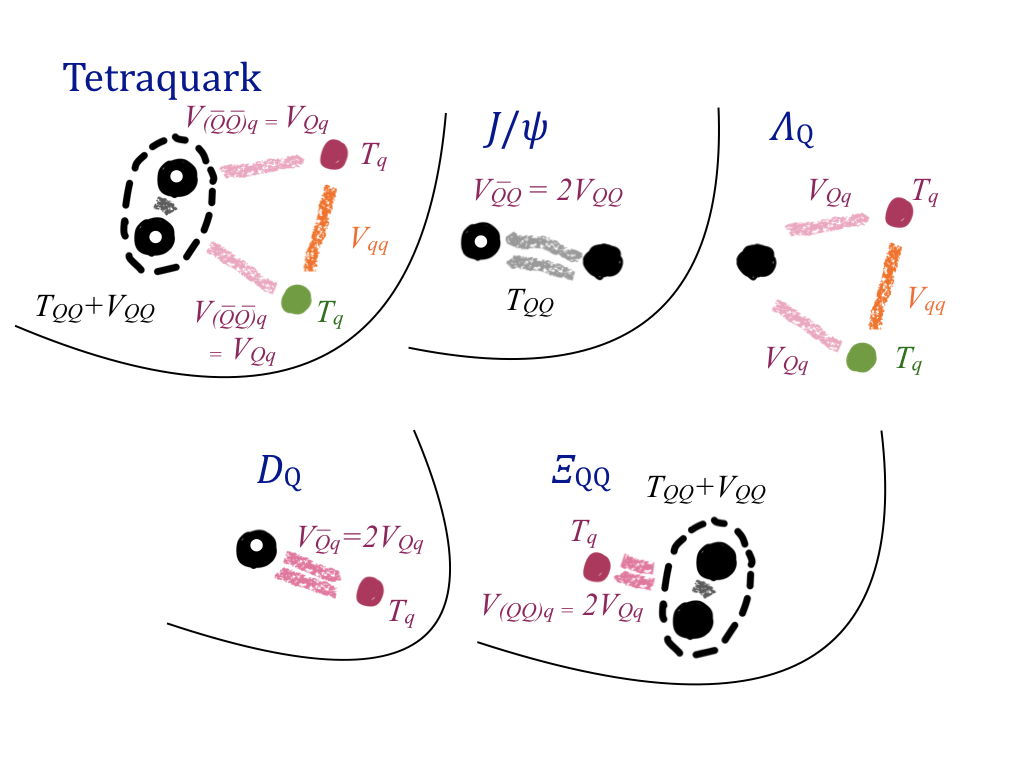} 
\vspace{-0.7cm}
\caption{Schematic illustration of multiquark states. First row shows states
taken into account by Lipkin \cite{Lipkin:1986dw}, second row shows two more states used
additionally in this paper. One thick line joining quarks represents
interaction in color ${\bf 3}$ or $\bar{{\bf 3}}$, whereas double line corresponds to color
singlet.}
\label{fig:Lipkin}
\end{figure}

To the best of our knowledge the first phenomenological attempt to estimate doubly heavy $QQ$
tetraquark mass was published by Lipkin in 1986 \cite{Lipkin:1986dw}
(although the fourfold heavy tetraquarks were discussed even earlier in 1982 \cite{Ader:1981db}). 
He used a variational method in the nonrelativistic quark model and one, rather
natural assumption, that light quarks see heavy (anti)quark pair as a single
object. Lipkin tried to use experimental data available at that time to
derive (almost) model-independent estimate of the tetraquark mass. This is
schematically shown in Fig.~\ref{fig:Lipkin} and leads to the following mass
formulae for tetraquark ($\mathcal{T}_{{Q}{Q}}$), $J/\Psi$ and $%
\Lambda_Q$: 
\begin{align}
M_{\mathcal{T}_{QQ}}&=2M+2m+T_{QQ}+V_{QQ}+2V_{Qq}+2T_q+V_{qq}, 
\notag \\
M_{J/\Psi}&=2m_Q+T_{QQ}+2V_{QQ},  \notag \\
M_{\Lambda_Q}&=m_Q+2m_q+2T_{q}+2V_{Qq}+V_{qq}  \label{Lipkin}
\end{align}
where for simplicity we have suppressed \emph{bars} over heavy quark symbol $%
Q$. 
Notation confronted with Fig.~\ref{fig:Lipkin} is self-explanatory. Note that
quark-quark interaction in color $\bar{{\bf 3}}$ is two times
weaker than antiquark-quark in ${\bf 1}$.

Formulae (\ref{Lipkin}) lead to the following upper bound for the tetraquark
mass: 
\begin{equation}
M_{\mathcal{T}_{{Q}{Q}}} \le M_{\Lambda_Q} +
\frac{1}{2} 
M_{J/\Psi}+
\frac{1}{2}
 \langle T_{QQ} \rangle.  \label{Lipkin_estimate}
\end{equation}
With the data available in 1986 one could not eliminate the
unknown average 
$\langle T_{QQ} \rangle$.
Nevertheless an important lesson can be drawn from Eq.~(\ref{Lipkin_estimate}%
) when plugging in numerical values with
 a spin averaged mass
$M_{\bar{c}c}=( 3M_{J/\Psi}+M_{\eta_c})/4$
rather than $M_{J/\Psi}$:
\begin{align}
M_{\mathcal{T}_{{c}{c}}}-M_{\mathrm{thr}}^c \le -55 ~\mathrm{%
MeV}+
\frac{1}{2} 
\langle T_{cc} \rangle  \label{charm}
\end{align}
with 
$M_{\mathrm{thr}}^c =M_D+M_{D^*}$.
We see from Eq.~(\ref%
{charm}) that the very existence of the bound heavy tetraquark
depends on the subtle balance between $-55$ MeV and $\langle T_{cc} \rangle$.
It is easy to convince oneself that adding the
D meson 
does not eliminate $\langle T_{cc} \rangle$.
Today we can repeat the same estimate for the $b$ case:
\begin{equation}
M_{\mathcal{T}_{{b}{b}}}-M_{\mathrm{thr}}^{b}\leq -262~\mathrm{MeV}+%
\frac{1}{2}
\langle T_{bb}\rangle .
\label{bottom}
\end{equation}%

Now, with the discovery of $\Xi_{cc}$ \cite{Aaij:2017ueg} one
can form a linear combination where the
troublesome $\langle T_{QQ} \rangle$ term drops out \cite{Gelman:2002wf}. To this end we have 
\begin{align}
M_{D_Q} & = m_Q+m_q+T_q+2 V_{Qq},  \notag \\
M_{\Xi_{QQ}} & =2 m_Q+m_q+T_{QQ}+V_{QQ}+T_q+2V_{Qq}.
\end{align}
With this new input we have one new relation: 
\begin{align}
M_{\mathcal{T}_{{Q}{Q}}} & \le M_{\Xi_{QQ}}+M_{\Lambda_Q}-M_{D_Q}.
\end{align}
where again we use $M_{D_{c}} =\left( 3M_{D^{\ast }}+M_{D}\right) /4$ (for a more accurate
choice see \cite{Gelman:2002wf}) and obtain numerically 
\begin{equation}
M_{\mathcal{T}_{cc}}\leq 3935~\mathrm{MeV},
\end{equation}%
which is 60 MeV above the threshold (as {\em e.g} in \cite{Ebert:2007rn}). This implies that, if
the inequality (\ref{Lipkin_estimate}) were saturated, $\langle T_{cc}\rangle \sim
230$~MeV. Since $m_{b}/m_{c}\sim 3$, one can reasonably assume that  $\langle
T_{bb}\rangle \sim \langle T_{cc}\rangle /3$ and we arrive at a conclusion
\begin{equation}
M_{\mathcal{T}_{{b}{b}}}-M_{\mathrm{thr}}^{b}\sim -224\;{\rm MeV}
\label{bottomnum}
\end{equation}
in a surprising agreement with much sophisticated quark model of Ref.~\cite{Karliner:2017qjm}.
In the literature one finds predictions for (\ref{bottomnum}) ranging from $-60$ \cite{Bicudo:2016ooe}
through $-100$ \cite{Ebert:2007rn}, $-120$ \cite{Eichten:2017ffp,Park:2018wjk} to $-400$ MeV~\cite{Du:2012wp}
(see Tab.~V in \cite{Ebert:2007rn} for other predictions).
Our simple analysis confirms 
that the binding of a putative doubly heavy $QQ$
tetraquark increases with increasing $m_Q$. This is true also in the case when the structure of a heavy diquark
can be resolved by the light quarks and repulsive ${\bf 6}$ channel is included \cite{Czarnecki:2017vco}.

\section{Soliton model for tetraquarks}

While the nonrelativistic approach may be applicable to heavy quarks, its credibility for the light sector
is certainly not at the same footing. In recent papers \cite{Yang:2016qdz} 
we have proposed a mean-filed description of heavy
baryons as a light quark-soliton and a heavy quark, where the soliton is
constructed from $N_{c}-1$ rather $N_{c}$ quarks. 
The model passes phenomenological tests. 

In the  limit $m_q \rightarrow \infty$ the soliton should not be sensitive to the properties of 
an object it is interacting with. Moreover, since the ground state soliton is in this case in flavour antitriplet
of spin zero \cite{Yang:2016qdz}, the spin of the heavy "nucleus" is irrelevant. Mass formula for the nonstrange
heavy baryons in flavor antitriplet is very simple
\begin{align}
M_{{\rm baryon}}^{Q,\boldsymbol{\overline{3}}}  &  =M_{\mathbf{\overline{3}}}^{Q}+\frac{2}{3}\delta_{\boldsymbol{\overline{3}}}
\end{align}
where $\delta_{\boldsymbol{\overline{3}}}$ can be extracted from the light hyperon spectrum, and $M_{\mathbf{\overline{3}}}$
is an average mass of the $\boldsymbol{\overline{3}}$ flavor multiplet including heavy quark mass $m_Q$, classical soliton mass
$M_{\rm sol}$
and soliton rotational energy \cite{Yang:2016qdz}. For strange heavy baryons the coefficient in front of $\delta_{\boldsymbol{\overline{3}}}$ 
is equal $-1/3$.
In \cite{Yang:2016qdz}  we never needed $m_Q$ and $M_{\rm sol}$ separately.
For nonstrange tetraquarks we therefore naturally have
\begin{align}
M_{{\rm baryon}}^{Q,\boldsymbol{\overline{3}}}  &  =M_{\mathbf{\overline{3}}}^{Q}+\frac{2}{3}\delta_{\boldsymbol{\overline{3}}}
+(m_{\bar{Q}\bar{Q}}-m_{Q})
\end{align}
and it is clear that now we need not only $m_{Q}$ but also 
$m_{\bar{Q}\bar{Q}}$. 
For a rough estimate
we can approximate $m_{\bar{Q}\bar{Q}}-m_{Q}\sim m_Q$, 
or we may assume a few percent binding
following {\em e.g.} Ref.~\cite{KarRos}.

In order to estimate effective $m_Q$ we first observe that differences of mean multiplet values, 
both for flavor $\boldsymbol{\overline{3}}$
and ${\bf 6}$, for bottom and charm, should be equal: 
$m_{b}-m_{c}=M_{\mathbf{\overline{3}}}^{b}%
-M_{\mathbf{\overline{3}}}^{c}=M_{\mathbf{6}}^{b}-M_{\mathbf{6}}^{c}$. Numerically we have:
\begin{align}
M_{\mathbf{\overline{3}}}^{b}%
-M_{\mathbf{\overline{3}}}^{c}=3327\;\text{MeV,} ~~&
M_{\mathbf{6}}^{b}-M_{\mathbf{6}}^{c}%
=3326\;\text{MeV}\label{eq:cbdiff}%
\end{align}
with $M_{\mathbf{\overline{3}},\mathbf{6}}^{Q}$ 
taken from \cite{Yang:2016qdz}.
We see perfect agreement between both flavor multiplets. Another piece of information comes from the hyperfine splittings
in $\mathbf{6}$ that are inversely proportional to the quark masses and have been estimated in~\cite{Yang:2016qdz}
yeilding:
\begin{equation}
{m_{c}}/{m_{b}}=0.29-0.31, \label{eq:cbratio}%
\end{equation}
which is compatible with the ratio obtained from the PDG \cite{PDG}.
Now, from Eqs.~(\ref{eq:cbdiff}) and (\ref{eq:cbratio}) we can determine
absolute effective masses 
\begin{align}
m_{c}   =1357\div1495\;\text{MeV,}~~&~~
m_{b}   =4685\div4821\;\text{MeV}. \label{eq:mQbaryon}%
\end{align}
The uncertainty 
in (\ref{eq:mQbaryon})
 is due to the uncertainty in the ratio (\ref{eq:cbratio}).  Masses (\ref{eq:mQbaryon}) are lower than
masses extracted from meson spectra: $m_c=1643$ and $m_{b}=4979$~MeV, which are 
compatible
with {\em e.g.} \cite{KarRos}.

\begin{figure}[h]
\centering
\includegraphics[width=6.2cm]{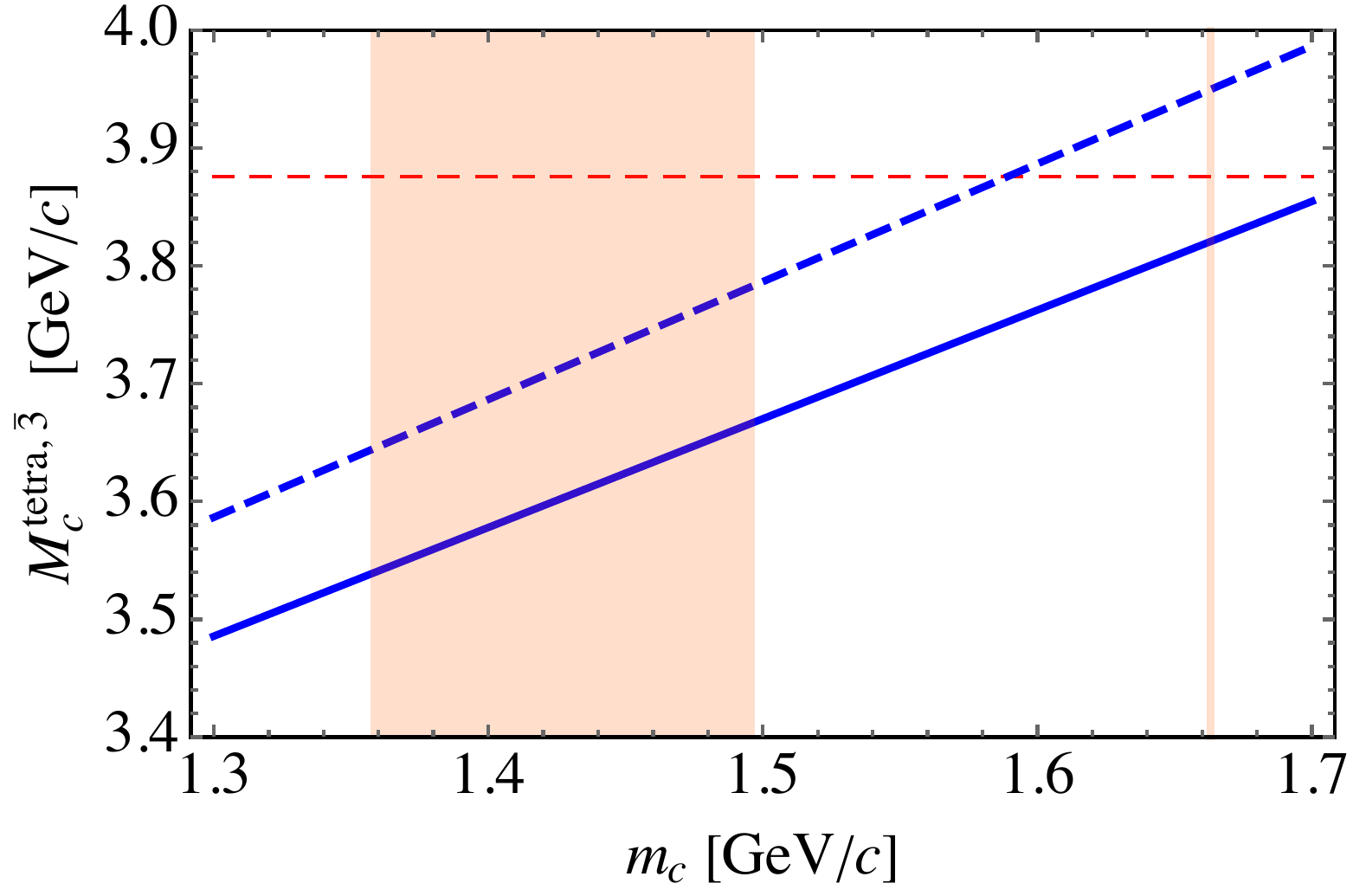}
\includegraphics[width=6.2cm]{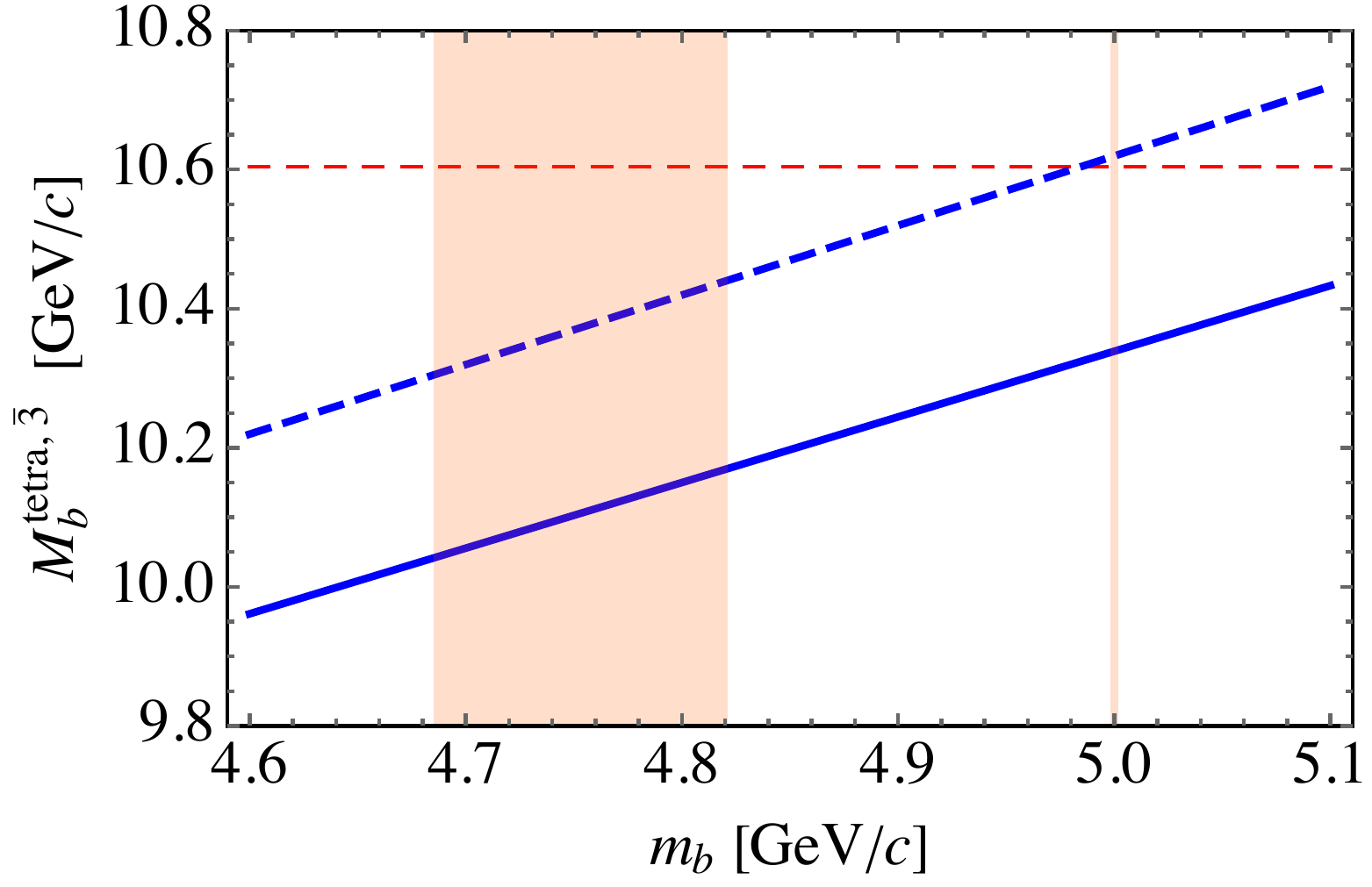} 
\caption{The lightest
$QQ$ tetraquark mass (charm -- left and bottom -- right)
as a function of $m_Q$ 
with (solid) and without (dashed) $\bar{Q}\bar{Q}$
binding contribution. Thin horizontal dashed (red) line corresponds to the
$D\,D^{\ast}$ or $B\,B^{\ast}$ threshold. Shaded areas indicate the heavy
quark mass range (\ref{eq:mQbaryon}). Solid vertical
line shows the heavy quark mass from Ref.~\cite{KarRos}.}%
\label{fig:3barmass}%
\end{figure}

We see from Fig.~\ref{fig:3barmass} that for heavy quark masses in the range of 
Eq.~(\ref{eq:mQbaryon}) 
both $cc$ and  $bb$ tetraquarks are rather deeply bound.
 For larger $m_Q$, compatible with mesonic spectra, $cc$ tetraquark is most likely
 unbound and $bb$ is most likely bound.
 
 \section{Summary}
 We have recalled arguments that $QQ$ tetraquarks are bound in the limit $m_Q \rightarrow \infty$ and 
 analyzed the mass spectrum of $cc$ and $bb$ tetraquarks with the help of the variational approach
 of Lipkin \cite{Lipkin:1986dw}. We than employed the quark-soliton model describing light degrees
 of freedom in the limit $N_c \rightarrow \infty$ used previously for heavy baryons with one heavy quark~\cite{Yang:2016qdz}
 to the problem of $QQ$ tetraquarks. We have argued that the light soliton does not distinguish the nature of the color
 ${\bf 3}$ heavy source, so that  heavy quark can be replaced by a heavy anti-diquark leaving the soliton unaffected.
 Unfortunately the anti-diquark properties have not been calculated within the soliton model. In fact in the present model
 the diquark should be considered an $N_c-1$ heavy quark system to neutralize the color of the soliton for $N_c>3$. This
 color structure (discussed briefly in \cite{Gelman:2002wf}) is completely different from the quark model picture where
 $QQ$ tetraquarks consist from two antiquarks and two quarks for any $N_c$. The "diquark" in the soliton approach is therefore
 amenable to an effective description, as the light sector that is represented by a soliton, and deserves further studies from
 this perspective. 
 
  \section*{Acknowledgments}
\noindent  This work was supported by the Polish NCN grant 2017/27/B/ST2/01314
and NAWA (Polish National Agency for Academic Exchange) Bekker program. The author
acknowledges stimulating discussions with H.-C. Kim and M.V. Polyakov and thanks Institute 
for Nuclear Theory at the University of Washington (where a part of this work has been completed) for
hospitality.


\begin{thebibliography}{99}

\bibitem{Aaij:2017ueg} 
  R.~Aaij {\it et al.} [LHCb Collaboration],
{\em  Phys.\ Rev.\ Lett.}\  {\bf 119}, 112001 (2017).
  
\bibitem{Mattson:2002vu} 
  M.~Mattson {\it et al.} [SELEX Collaboration],
  {\em Phys.\ Rev.\ Lett.\   }{\bf 89}, 112001 (2002).
  
\bibitem{Ocherashvili:2004hi} 
  A.~Ocherashvili {\it et al.} [SELEX Collaboration],
 {\em  Phys.\ Lett.\ B} {\bf 628}, 18 (2005).
  
\bibitem{PDG}
M. Tanabashi {\em et al.} (Particle Data Group), {\em Phys. Rev. \ D} {\bf 98}, 030001 (2018).


\bibitem{Gelman:2002wf} 
  B.~A.~Gelman and S.~Nussinov,
{\rm  Phys.\ Lett.\ B} {\bf 551}, 296 (2003).
  
\bibitem{Manohar:1992nd}
  A.~V.~Manohar and M.~B.~Wise,
 {\em  Nucl.\ Phys.\  B} {\bf 399}, 17 (1993).
  
\bibitem{Cohen:2006jg} 
  T.~D.~Cohen and P.~M.~Hohler,
  {\em Phys.\ Rev.\  D} {\bf 74}, 094003 (2006).
  
\bibitem{Cai:2019orb} 
  Y.~Cai and T.~Cohen,
  arXiv:1901.05473 [hep-ph].
 
  
\bibitem{Lipkin:1986dw}  H.~J.~Lipkin,  
{\em Phys.\ Lett.\  B} {\bf 172}, 241 (1986).

\bibitem{Ader:1981db}
  J.~P.~Ader, J.~M.~Richard and P.~Taxil,
  {\em Phys.\ Rev.\ D} {\bf 25},  2370 (1982).

\bibitem{Ebert:2007rn} 
  D.~Ebert, R.~N.~Faustov, V.~O.~Galkin and W.~Lucha,
{\em  Phys.\ Rev.\ D} {\bf 76}, 114015 (2007).

\bibitem{Karliner:2017qjm} 
  M.~Karliner and J.~L.~Rosner,
  {\em Phys.\ Rev.\ Lett.\ }{\bf 119}, 202001 (2017).
  
\bibitem{Bicudo:2016ooe} 
  P.~Bicudo, J.~Scheunert and M.~Wagner,
  {\em Phys.\ Rev.\ D} {\bf 95}, 034502 (2017).
  
\bibitem{Eichten:2017ffp} 
  E.~J.~Eichten and C.~Quigg,
 {\em  Phys.\ Rev.\ Lett.\ }  {\bf 119}, 202002 (2017).
 
  
\bibitem{Park:2018wjk} 
  W.~Park, S.~Noh and S.~H.~Lee,
 {\em  Nucl.\ Phys.\ A} {\bf 983}, 1 (2019).
  
\bibitem{Du:2012wp} 
  M.~L.~Du {\em et al.},
{\em   Phys.\ Rev.\ D }{\bf 87},  014003 (2013).
  
\bibitem{Czarnecki:2017vco} 
  A.~Czarnecki, B.~Leng and M.~B.~Voloshin,
 {\em  Phys.\ Lett.\ B} {\bf 778}, 233 (2018).
  
\bibitem{Yang:2016qdz} 
  G.~S.~Yang, H.~C.~Kim, M.~V.~Polyakov and M.~Praszalowicz,
{\em   Phys.\ Rev.\ D }{\bf 94}, 071502 (2016) and
 {\em Phys.\ Rev.\ D }
{\bf 96}, 094021 (2017),
  Erratum: [{\em Phys.\ Rev.\ D} {\bf 97}, 039901 (2018)].

  
\bibitem{KarRos}
M.~Karliner and J.~L.~Rosner, {\em Phys. Rev. D} {\bf 90}, 094007 (2014).
  
\end{thebibliography}
\end{document}